\documentclass[%
 reprint,
prb,
]{revtex4-2}

\usepackage{physics}
\usepackage{color,soul}
\usepackage{dcolumn}% Align table columns on decimal point

\usepackage{amsmath}
\usepackage{float}
\usepackage{graphicx}
\usepackage{siunitx}
\usepackage{bm}

\usepackage{dcolumn}% Align table columns on decimal point
\usepackage{amsmath}
\usepackage{float}
\usepackage{graphicx}
\usepackage{siunitx}
\usepackage[colorlinks=true, allcolors=blue]{hyperref}
\begin{document}

%\preprint{APS/123-QED}
\title{Inverse Design of High-NA Metalens for Maskless Lithography}

%%%% PRB style
\author{Haejun Chung*}
\affiliation{Department of Electronic Engineering, Hanyang University, Seoul, 04763, South Korea, e-mail: haejun@hanyang.ac.kr}

\author{Feng Zhang}%
\affiliation{LumArray, Inc., 15 Ward Street, Somerville, MA 02143, USA}

\author{Hao Li}
\affiliation{Department of Applied Physics and Energy Sciences Institute, Yale University, New Haven, Connecticut 06511, USA}
\author{Owen D. Miller}
\affiliation{Department of Applied Physics and Energy Sciences Institute, Yale University, New Haven, Connecticut 06511, USA, e-mail: owen.miller@yale.edu}
\author{Henry I. Smith*}%
\affiliation{LumArray, Inc., 15 Ward Street, Somerville, MA 02143, USA}
\affiliation{Department of Electrical Engineering and Computer Science, Massachusetts Institute of Technology (MIT), Cambridge, MA 02139 USA, e-mail: hismith@mit.edu}

%%% Optica style
%\author[1]{Haejun Chung}
%\author[1,*]{Owen D. Miller}
%\affil[1]{Department of Applied Physics and Energy Sciences Institute, Yale University, New Haven, Connecticut 06511, USA}
%\affil[*]{Corresponding author: owen.miller@yale.edu}

% To be edited by editor
% \dates{Compiled \today}

%\ociscodes{(140.3490) Lasers, distributed feedback; (060.2420) Fibers, polarization-maintaining; (060.3735) Fiber Bragg gratings.}

% To be edited by editor
% \doi{\url{http://dx.doi.org/10.1364/optica.XX.XXXXXX}}

\begin{abstract}
We demonstrate an axisymmetric inverse-designed metalens to improve the performance of zone-plate-array lithography (ZPAL), one of the maskless lithography approaches, that offer a new paradigm for nanoscale research and industry. First, we derive a computational upper bound for a unit-cell-based axisymmetric metalens. Then, we demonstrate a fabrication-compatible inverse-designed metalens with 85.50\% transmission normalized focusing efficiency at 0.6 numerical aperture at 405nm wavelength; a higher efficiency than a theoretical gradient index lens design (79.98\%). We also demonstrate experimental validation for our axisymmetric inverse-designed metalens via electron beam lithography. Metalens-based maskless lithography may open a new way of achieving low-cost, large-area nanofabrication.
\end{abstract}

%\setboolean{displaycopyright}{true}

\maketitle

\section{Introduction}
Zone-plate-array lithography (ZPAL) creates patterns of arbitrary geometry by overlapping focal spots produced by an array of microlenses~\cite{smith2006zone,chao2005immersion,smith1996proposal}. Such maskless lithography avoids the cost and delay time associated with procuring a photomask~\cite{pang2006inverse}. It also avoids troublesome coherence problems associated with mask-based projection photolithography. Figure~\ref{schematic}(a) is a schematic of the ZPAL technology. A significant obstacle for ZPAL is the need for a high-efficiency, high-numerical-aperture (NA) microlenses. Recently, the advent of the "metalens" has led to the development of structured, wavelength-scale devices offering superior performance to conventional approaches~\cite{wang2016chromatic,khorasaninejad2016polarization,khorasaninejad2016metalenses,lalanne2017metalenses,shrestha2018broadband,chen2018broadband}. A standard metalens design approach, called unit-cell design, stitches subwavelength resonators together into a larger device, demonstrating the possibility of focusing~\cite{khorasaninejad2016polarization,khorasaninejad2016metalenses,lalanne2017metalenses,shrestha2018broadband,chen2018broadband}, but it has suffered from low focusing efficiencies or low-NA restrictions. Recently, it has been shown that the unit-cell design approach cannot have high efficiency at high-NA due to a fast oscillation of the field phase and amplitude profiles and a local periodicity violation~\cite{chung2020high}.

In this work, we apply inverse design, a large-scale computational technique~\cite{miller2012photonic, su2017inverse, lin2019topology, molesky2018inverse, chung2020tunable, chung2021inverse}, to discover a high-NA metalens at 405nm wavelength to realize a high-resolution, high-efficiency ZPAL lithography system. 
Inverse design enables fast computation of Figure-of-Merit (FoM) gradients with respect to design parameters over the entire design space. Fabrication-constraints are implemented by penalization technique~\cite{stolpe2001trajectories, neves2002topology} and constrained optimization~\cite{reuther1999constrained, bertsekas2014constrained}, enabling fast discovery of fabrication-compatible designs.

\section{Method}
\subsection{Zone-plate-array lithography: The microlenses}

\begin{figure*}[ht]
\centering
\includegraphics[width=1.0\linewidth]{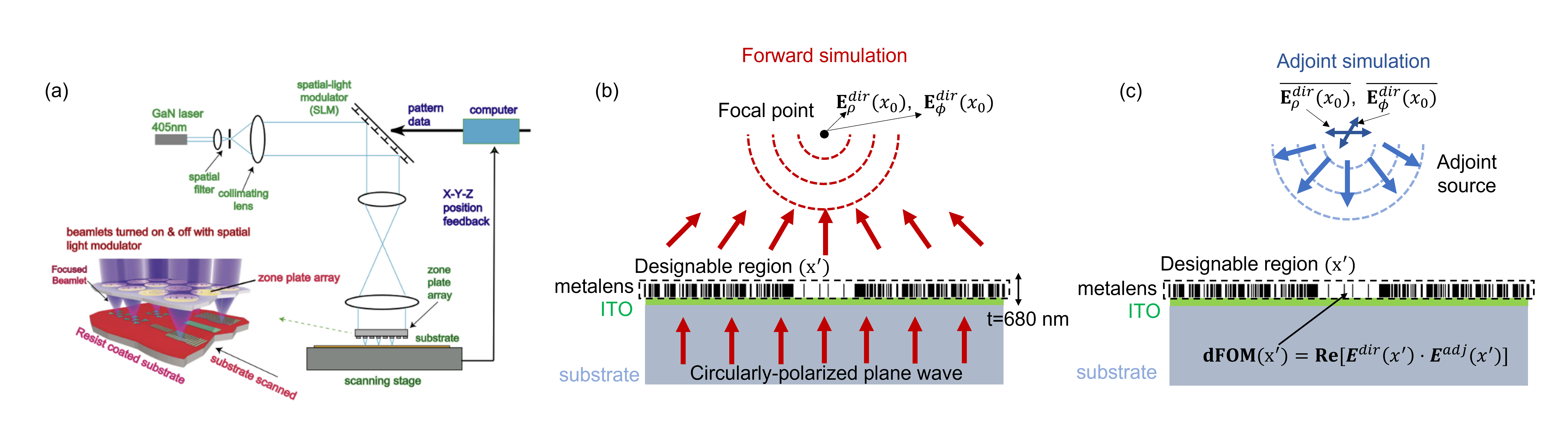}
\caption{(a) Schematic depiction of zone-plate-array lithography (ZPAL). A CW laser illuminates a spatial-light modulator, each pixel of which controls the light intensity to one zone-plate of the array. By adjusting the focal-spot intensity from zero to maximum in a quasi-continuous manner, linewidth can be controlled and proximity-effects corrected. By moving the stage under computer control, while intelligently modulating focal-spot intensities, patterns of arbitrary geometry can be written. (b) Schematic depiction of forward simulation of inverse design. A circularly-polarized plane wave is excited from the substrate. $E^{dir}_\rho(x_0), E^{dir}_\phi(x_0)$ at the focal point are stored for the adjoint simulation. (c) A conjugated $E^{dir}_\rho(x_0), E^{dir}_\phi(x_0)$ field now back-propagates to obtain adjoint electric fields at the designable region ($x'$).}
\label{schematic}
\end{figure*}

% Zone-plate-array lithography (ZPAL) creates patterns of arbitrary geometry by overlapping focal spots produced by an array of microlenses. Such a maskless lithography  avoids the cost and delay time associated with procuring a photomask, but it also avoids troublesome coherence problems associated with mask-based projection photolithography. Figure~\ref{schematic}(a) is a schematic representation of the ZPAL technology. 
% \subsubsection{The microlenses}

	The original implementation of ZPAL utilized binary $\pi$-phase-shifting zone plates formed in a spin-on glass with index of refraction of n $=$ 1.47. It is well known that binary $\pi$-phase gratings~\cite{li2012photonic}, in which the zeroth order is canceled and second and third orders are cut off,  have a diffraction efficiency of $\sim$42\% into the $+1$-order~\cite{menon2006experimental}. Based on this, one would expect a maximum focal efficiency of about 40\% for a binary $\pi$-phase-shifting zone plate~\cite{menon2006experimental}. Assuming this, 60\% of the transmitted light would constitute background exposure. This background is composed of residual zeroth order as well as a -1-order virtual focus and third order foci, both real and virtual. Because ZPAL’s writing scheme consists of overlapping non-coherent focal spots, interference effect are avoided and linear superposition applies. As a consequence, proximity-effect correction (PEC) can be easily implemented if the focal spot’s point-spread function, including the broad background, are known. Although PEC can compensate for a high background level (as is done in scanning-electron-beam lithography where electron backscattering produces a large, broad background) it nevertheless is highly desirable to reduce this background.  

In projection photolithography, it is common to express the minimum practical linewidth achievable in dense patterns as,

\begin{equation}
\textrm{W}_\textrm{min} = \textrm{k}_1\lambda/\textrm{NA}
\label{eq1}
\end{equation}
where $\lambda$ is the optical wavelength, NA is the len’s numerical aperture and $\textrm{k}_1$ is a proportionality factor, obtained empirically. In practice, the lower the background level, the lower one can push $\textrm{k}_1$ in pursuit of finer linewidths.  

In seeking microlenses with higher focal efficiencies than $\pi$-phase zone plates, the obvious first approach would be shaping the individual zones, to produce a gradual phase delay across them, a scheme referred to as “blazing” in the case of gratings~\cite{wilson2003recent, loewen1977grating}. Such a microlens is commonly called a Fresnel lens~\cite{miyamoto1961phase, ceglio1978micro}. Aside from the difficulty of performing such blazing with existing nanofabrication tools, it is well known that the focal efficiency of Fresnel lenses drops rapidly with increasing NA~\cite{lalanne2017metalenses}. Some improvement can be achieved by so-called effective-index modulation (EIM)~\cite{hassan2020integrated, alexopoulos2010effective} in which the thickness of the dielectric is uniform from zone to zone, but across each zone dielectric material is selectively removed to achieve a desired variation of phase delay. Not surprisingly this approach also encounters decreasing focal efficiency with increasing NA~\cite{chung2020high}. Our intuitive models about how light behaves in transiting and emerging from such complex structures break down when diffraction angles get large. 

In recent years, several groups have investigated so-called metasurfaces which includes metalenses~\cite{chung2020high, khorasaninejad2016polarization, ni2013ultra, khorasaninejad2016metalenses, shrestha2018broadband, yoon2020single, yoon2021printable}. The term metasurface was introduced to convey the notion of metal or dielectric structures in which the dimensions of features are below the wavelength of light. In many cases, one can think of these small structures behaving as sub-wavelength antennas. 

  High-Numerical-Aperture (NA) metalenses offer the possibility of significantly improving the resolution of the ZPAL maskless lithography system. Here, we study the theoretical efficiency limits of high-NA metalenses and then use inverse design to realize a high-efficiency high-NA metalens. Our inverse-designed metalens shows 85\% efficiency, close to the theoretical limit. We also fabricated metalens on SiO$_2$ substrates to demonstrate the feasibility of our approach.

\subsection{Upper bound of unit-cell design}

\begin{figure*}[ht]
\centering
\includegraphics[width=1.0\linewidth]{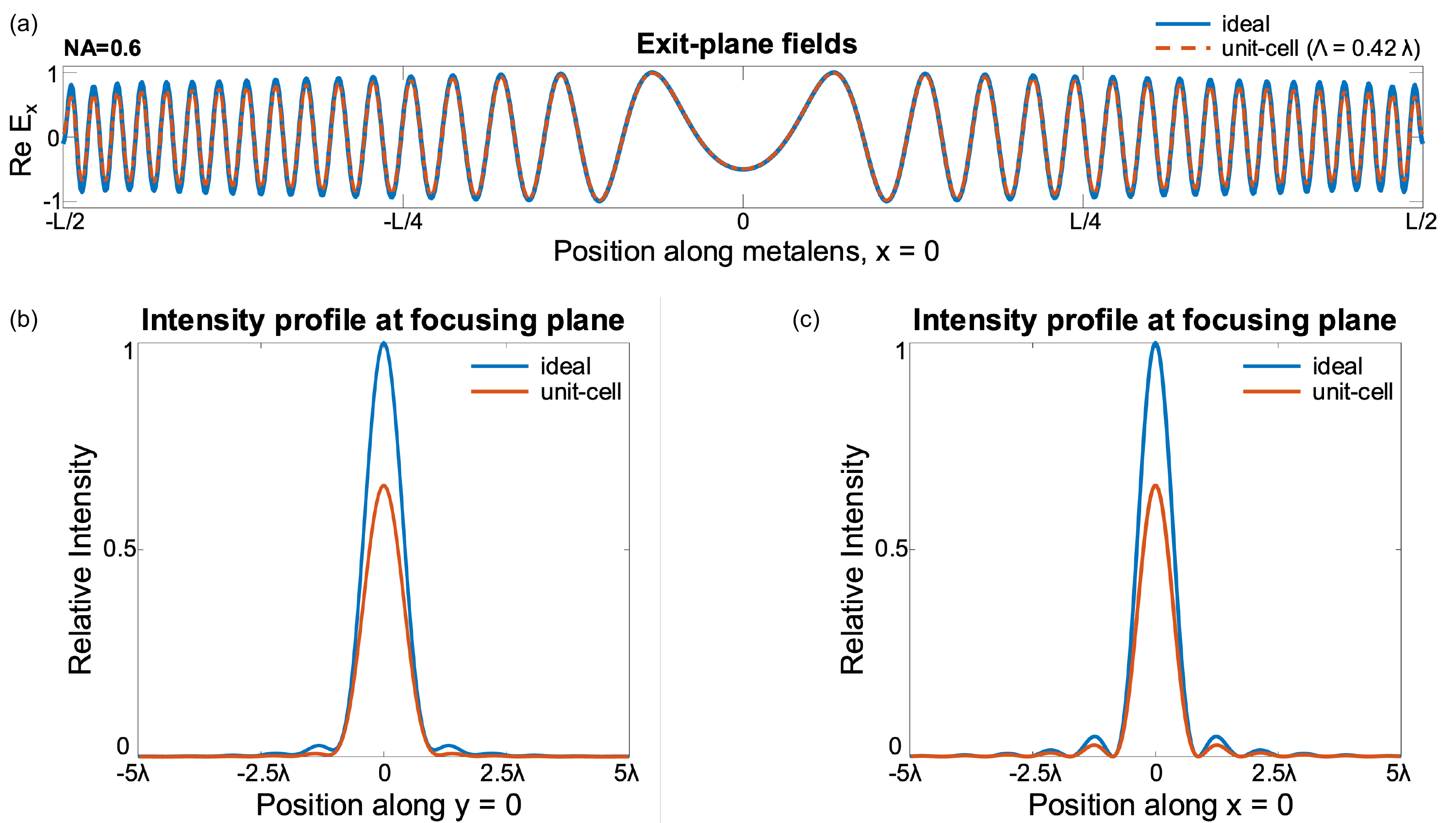}
\caption{(a) The exit fields of a unit cell design (red) cannot exactly match those of the ideal focusing metalens (blue). (b,c) These field differences at the exit planes lead to reduced intensities at the focal point, as plotted along both transverse directions.}
\label{theory}
\end{figure*}

To start, we ask: what is the largest efficiency a metalens designed by the unit-cell method could achieve? We apply an analytical technique first developed by some of the authors in Ref.~\cite{chung2020high}. The idea is as follows. In the unit-cell method, one approximates the response of each metalens unit as though it were part of a periodic structure. The collective plane of outgoing fields should then comprise the stitched-together fields from each individual unit cell. Yet even at this simple level of construction, one can already identify a source of error that must emerge. The stitched-together fields provably are not exact solutions of Maxwell's equations. The most favorable assumption is to take the outgoing fields to be the closest projection of the stitched-together fields onto a complete Maxwell basis. Once one makes this step, however, then it becomes clear that some efficiency must have been lost, as the projected fields will not propagate to the focal point identically to how the original fields would have. The resulting efficiency is an upper bound to the focusing efficiency of a unit-cell design. The mathematical derivations of the various statements above is given in Ref.~\cite{chung2020high}. We extend the 2D computations from that paper to 3D computations that are shown in Fig.~\ref{theory}. The exit plane fields show how the optimal stitched-together (``ideal'') fields are modified, just moderately, into the nearest projection of Maxwell fields. These differences can be more visibly significant in the focal plane, where, for parameters matching those of the later metalenses (NA = 0.6, unit-cell period = 0.42$\lambda$), the maximum focusing efficiency of a unit-cell design is about 63\%.

\subsection{Inverse design of metalenses}

\begin{figure*}[ht]
\centering
\includegraphics[width=1.0\linewidth]{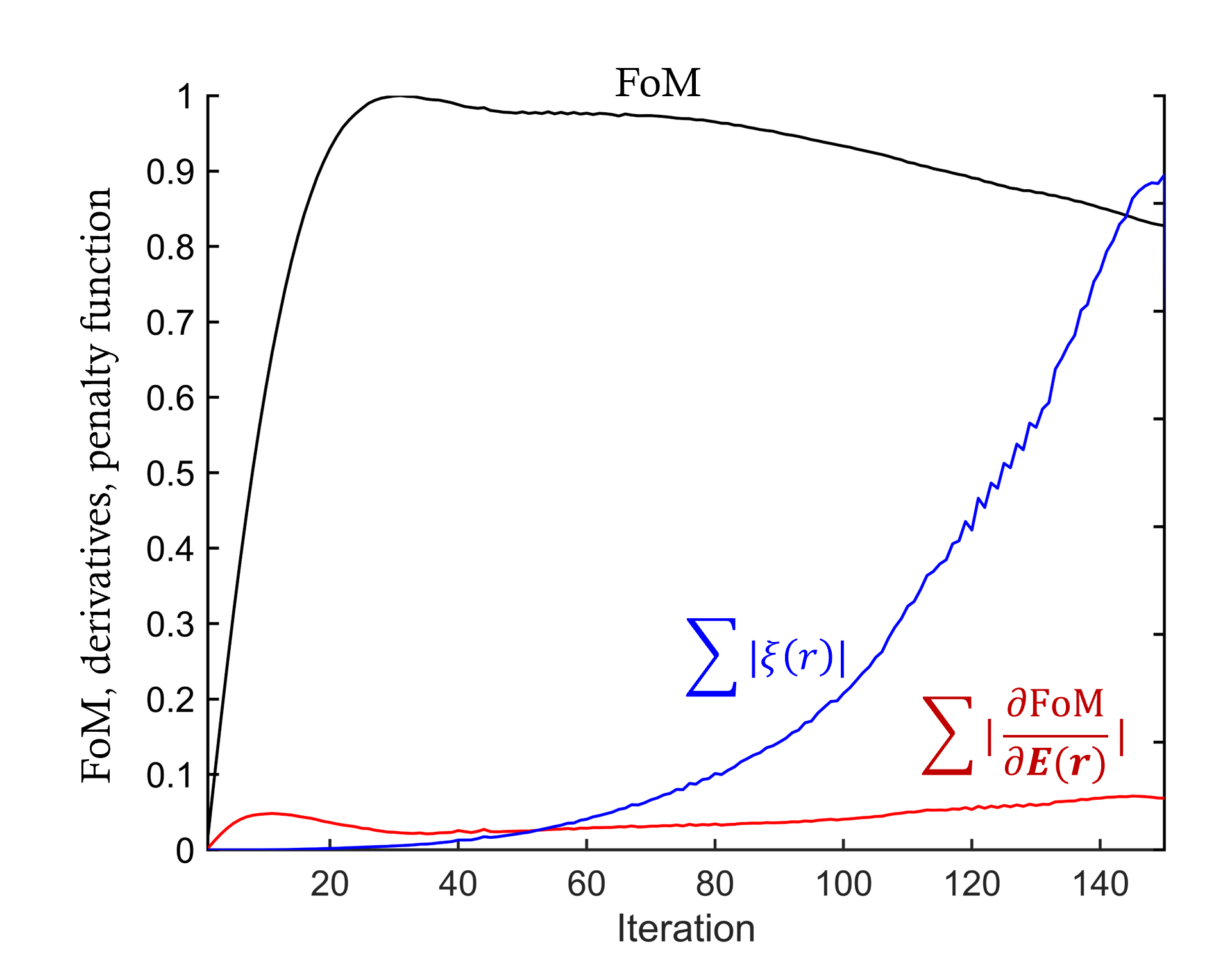}
\caption{Plot of the Figure-of-Merit (FoM), the sum of derivatives of FoM with respect to electric fields ($\sum|\frac{\partial \textrm{FoM}}{\partial \vb{E}(\vb{r})}|$), and the penalization factors ($\sum|\xi(\vb{r})|$) versus inverse design iterations. For the first 50 iterations, derivatives of FoM dominate over penalization factors, which means that the adjoint derivatives are the main driving factors for updating design parameters. After the 50$_\textrm{th}$ iteration, penalization factors gradually dominate the adjoint derivatives, making the design parameters converge to the binary level.}
\label{fom_over_iteration}
\end{figure*}

Inverse design, with “adjoint”-based optimization, computes gradients with respect to all structural degrees of freedom within forward and backward simulations~\cite{miller2012photonic, chung2020high, chung2022computational, lin2019topology}. Its origin can be traced back to circuit theory, control theory~\cite{Pontryagin1962}, quantum dynamics~\cite{Demiralp1993}, and deep learning~\cite{Werbos1994, lecun2015deep, Rumelhart1986}. It has been used in nanophotonics~\cite{piggott2015inverse, miller2012photonic, su2017inverse} for emerging applications such as tunable metasurfaces~\cite{chung2020tunable}, solar cells~\cite{lalau2013adjoint}, waveguide demultiplexers~\cite{su2017inverse,piggott2015inverse}, photolithographic mask~\cite{nam2022photolithographic}, and CMOS image sensors~\cite{catrysse2022subwavelength, zhao2021perfect}. Inverse design requires definition of a FoM and its design parameters. In this work, we confine our design space to axisymmetric geometries for faster computations; we can expect this assumption to have little to no effect on the ultimate focusing efficiency, since the ideal focusing functionality can be expected to exhibit axial symmetry as well. We use the full-wave Finite Difference Time Domain (FDTD) method~\cite{oskooi2010meep} to avoid assumptions, potentially ending up with focusing efficiency drops in unit-cell approaches~\cite{chung2020high}, in the design procedure. For the FOM, we use intensity maximization at the focal point, which generally corresponds to the maximization of focusing efficiency in the far field~\cite{shim2019maximal,Schab2022upper}. Our FoM is given by

\begin{equation}
\mathcal{F} = |\vb{E}(\textrm{x}_0)|^2
\label{eq2}
\end{equation}
where $\textrm{x}_0$ is a focal spot at the given lens dimension and NA, as shown in Fig.~\ref{schematic}(b), (c). Required computations can be further optimized through near-field to far-field transformation~\cite{lin2019topology} or planewave order decomposition~\cite{chung2020high}; however, the benefits of applying these techniques are marginal in high-NA metalenses due to a short focal length. Each geometrical parameter is a nano-ring with 40nm width, 860nm height. The radius of a nano-ring is defined as $40\cdot\textrm{N}$ nm (N is the index of geometry parameter). Our metalens design consists of a SiO$_2$ substrate (measured refractive index of 1.47 at 405 nm wavelength), 100-nm-thick ITO (measured refractive index of 2.10 at 405 nm wavelength), and 680-nm-thick electron-beam resist CSAR-62 (measured refractive index of 1.59 at 405 nm wavelength). For the geometrical degrees of freedom of nano-rings, we confine the density of CSAR-62 between 0 and 1 at every nano-ring (40-nm-width, 680-nm-height), and then add penalization functions~\cite{stolpe2001trajectories, neves2002topology} to the FoM in Eq.~\ref{eq2} to ensure a binary-material constraint. A circularly-
polarized plane wave is excited at the substrate, and then $E^{dir}_\rho(x_0), E^{dir}_\phi(x_0)$ at the focal point are calculated for the adjoint simulation. A conjugated $E^{dir}_\rho(x_0), E^{dir}_\phi(x_0)$ field back-propagates with an adjoint source given by

\begin{equation}
\vb{J}_{\rm adj} = -i\omega \vb{P}_{\rm adj} = -i\omega \partial \mathcal{F} / \partial \vb{E}
\label{eq3}
\end{equation}
where $\vb{P}$ is an source dipole density~\cite{miller2012photonic}. At the design space, derivatives of the FoM can be calculated via $\partial \mathcal{F} / \partial \varepsilon(\vb{x}) = \Re \left[ \vb{E}_{\rm dir}(\vb{x}) \cdot \vb{E}_{\rm adj}(\vb{x}) \right]$, where $\vb{E}_{\rm dir}$ is a direct electric field calculated via the forward simulation and $\vb{E}_{\rm adj}$ is an electric field obtained via the adjoint simulation~\cite{chung2020high, miller2012photonic}. The transmission normalized focusing efficiency, in this work, is defined as 

\begin{equation}
\eta = \frac{{\rm P}_{\rm 3rd,zero}}{{\rm P}_{\rm trans}}
\label{focusing_Eff}
\end{equation}
where ${\rm P}_{\rm 3rd,zero}$ is the electromagnetic power within the third zeroes of the Airy disk area while ${\rm P}_{\rm trans}$ denotes a transmitted electromagnetic power at the metalens surface. This definition of the transmission normalized focusing efficiency is different than that in an imaging application where focusing efficiency is generally defined as $\eta = \frac{{\rm P}}{{\rm P}_{\rm inc}}$. This difference arises from the nature of the lithography application, where the incidence power can be increased easily; thus, we can ignore reflected power, unlike imaging applications where the incoming light is relevant.

With our approach, a combination of forward and backward simulations takes approximately 330 seconds, including pre, and post-processing of adjoint derivatives on 64 cores in our workstation (AMD Ryzen Threadripper PRO 3995WX, 2.7GHz processors). The FoM rapidly increases then converges in about the 25th iteration, after which the penalization~\cite{stolpe2001trajectories, chung2020high} transforms the grayscale refractive indices to binary values (i.e., air or CSAR-62), which takes around another 120 iterations as shown in Fig.~\ref{fom_over_iteration}. The averaged penalization factor over the designable region ($\sum|\xi(\vb{r})|$) is set to a relative value compared to the sum of derivatives of FoM with respect to electric fields ($\sum|\frac{\partial \textrm{FoM}}{\partial \vb{E}(\vb{r})}|$). By taking relative penalization factors into account, the design parameters can be updated more systematically (e.g., adjoint derivatives dominant: 1st $\sim$ 40th iterations, penalization factor dominant: 40th $\sim$ 150th iterations). The penalization process can be further tuned with Gaussian filters~\cite{sell2017large}, density filters~\cite{wang2019robust, wang2011robust} or level-set
methods~\cite{vercruysse2019analytical, mansouree2019metasurface}. 

\section{Result}

\begin{figure*}
\centering
\includegraphics[width=1.0\linewidth]{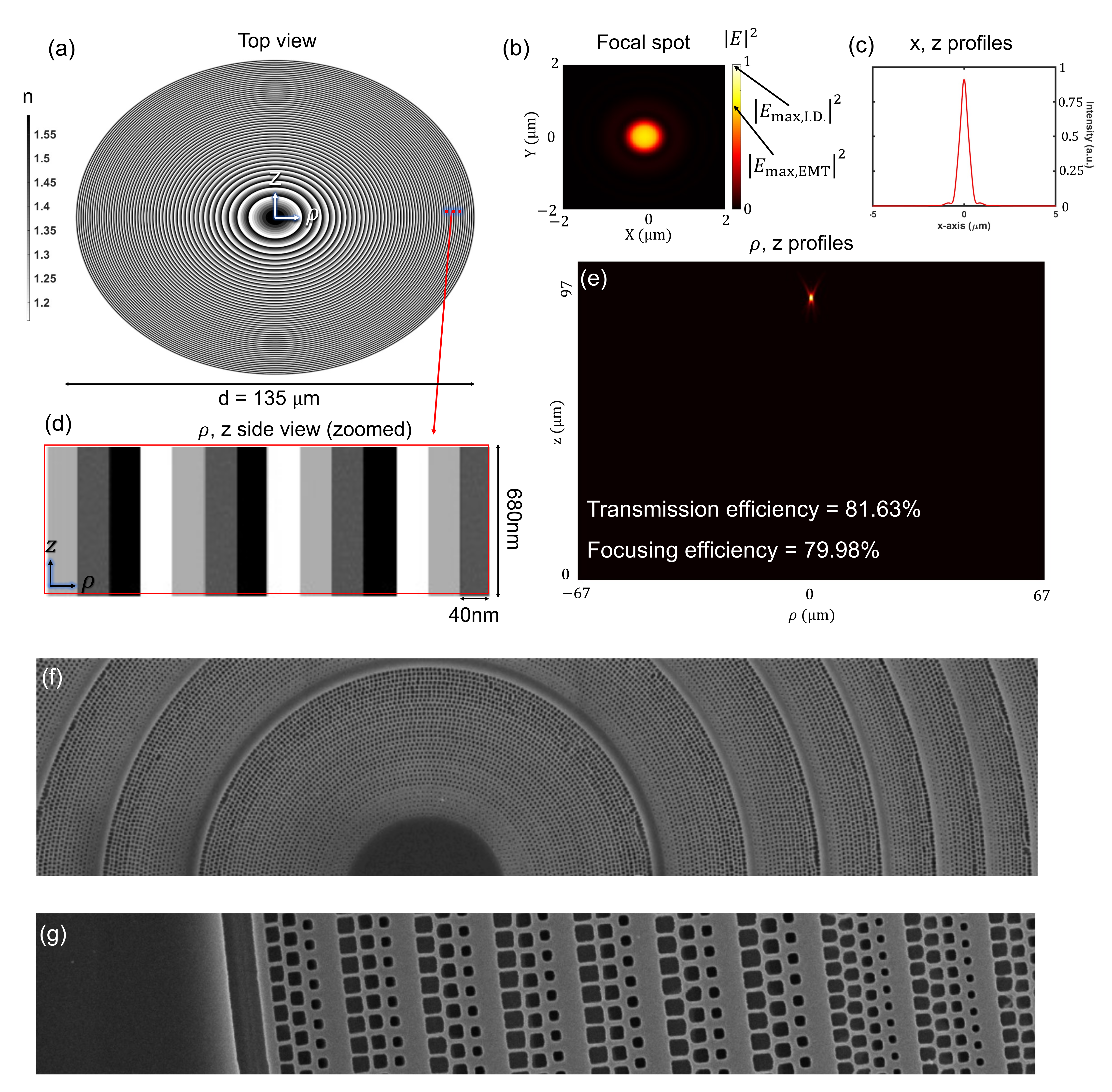}
\caption{Effective-Index Modulation (EIM)-based theoretical lens design. (a) Top view of the EIM-based lens for NA = 0.57 (f=97$\mu$m, d=135$\mu$m). (b) Focal spot intensity profile in the transverse (xy) plane at the focal distance of f = 97 $\mu$m. (c) A transverse cut of the focal spot intensity profile in simulation. The full width at half intensity maximum (FWHM) of the spot is 0.571 $\mu$m. (d) $\rho,z$ side view (zoomed) of the EIM-lens. Thickness is 680nm, covering 2-$\pi$ phase change with a maximum refractive variation of 1.0 to 1.59. Each pixel occupies 40nm radial space in cylindrical symmetry, corresponding to our fabrication resolution. (e) Longitudinal ($\rho$,z) intensity profile where transmission efficiency and transmission normalized focusing efficiency are 81.63\%, 79.98\%, respectively. (f) SEM of the EIM metalens which experimentally had a focal efficiency of about 48\%. (g) SEM of the outer zones of the EIM metalens.}
\label{EMT_lens}
\end{figure*}

In this section, we design and fabricate effective-medium-based and inverse-designed metalenses. Both approaches assume that we can have control within a fabrication constraint of 40nm (width) $\times$ 680nm (height) on the size of nano-ring. To explore a computational upper limit of the given design problem, the Effective Index Modulation (EIM) approach can have a material degree of freedom in grayscale (gradient refractive index) as shown in Fig.~\ref{EMT_lens}(a). The inverse design only allows a refractive index of 1.59 (CSAR-62) and 1.00 (air). The refractive indices of each nano-ring were chosen by matching a required phase profile $\Delta\phi(r)=\frac{2\pi}{\lambda}(f-\sqrt{r^2+f^2})$ and transmitted phase $\phi_{\rm trans.}(r)=\frac{2\pi {\rm n}(r) {\rm h}}{\lambda}$ of the gradient index material, where $\lambda$ is the design wavelength (405nm), r is the coordinate of each nano-ring, f is the focal length (97 um), and ${\rm n}(r)$ is a refractive index at the radius r. The minimum grid spacing of the simulations was enforced to 40nm to satisfy fabrication constraints, and then the final design was validated with a finer grid spacing to guarantee the convergence of the simulation. A circularly-polarized planewave was excited at the SiO$_2$ substrate, as shown in Fig.~\ref{schematic}(b). The incident power was calculated and then compared against a transmitted power at the plane 2$\lambda$ above the metalens to compute transmission efficiency shown in Fig.~\ref{EMT_lens}(e) (transmission normalized focusing efficiency is defined in Eq.~\ref{focusing_Eff}). We also verify robustness of our final design by simulating fabrication imperfections, and maintain relatively high efficiency for moderate fabrication imperfections as shown in Fig.~\ref{Robustness_check} (Appendix).

\subsection{Effective Index Modulated (EIM) Metalens}

\begin{figure*}
\centering
\includegraphics[width=1.0\linewidth]{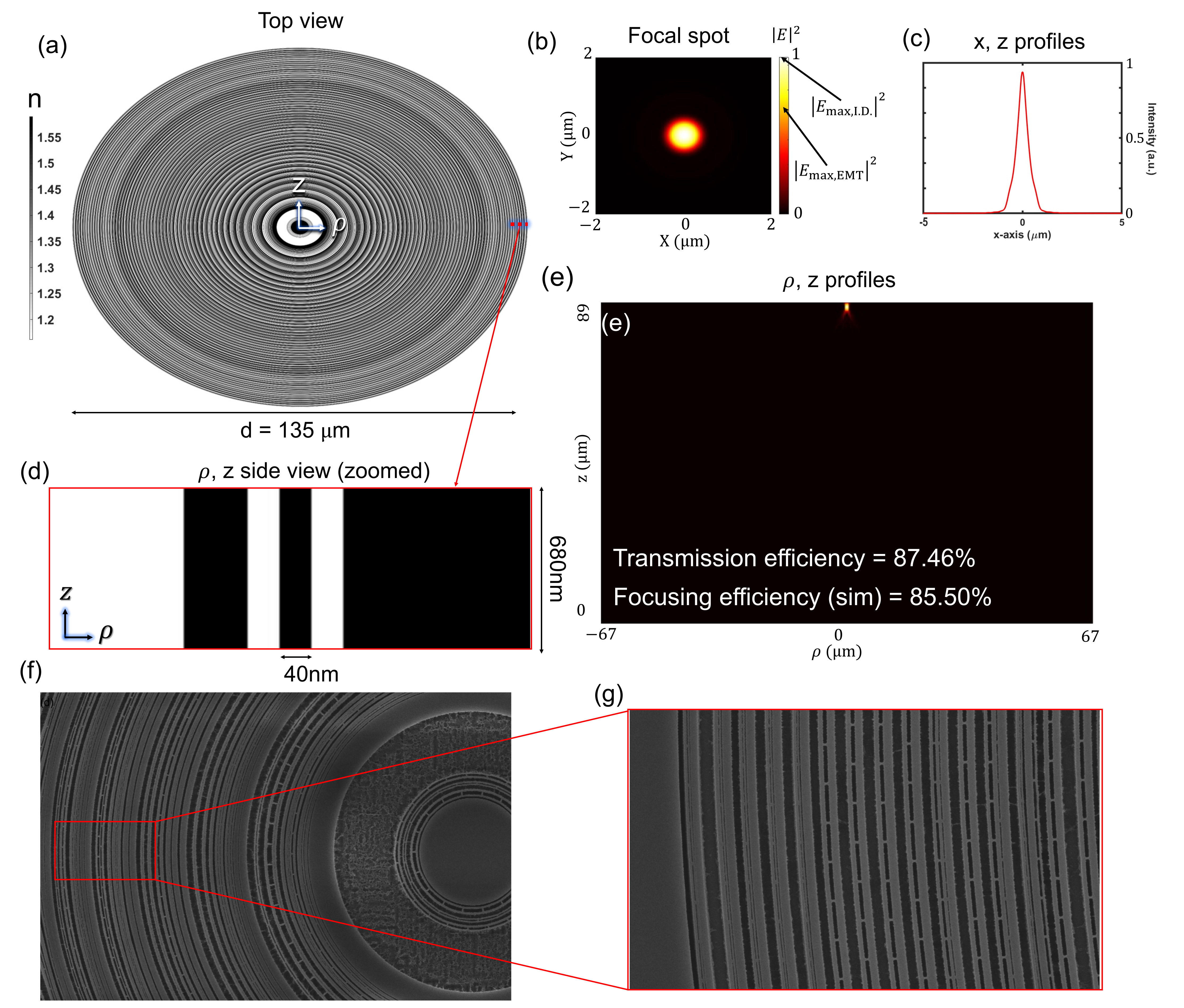}
\caption{Inverse-designed metalens. (a) Top view of the inverse-designed metalens for NA = 0.60 (f=89$\mu$m, d=135$\mu$m). (b) Focal spot intensity profile in the transverse (xy) plane at the focal distance of f = 89 $\mu$m. (c) A transverse cut of the focal spot intensity profile in simulation. The full width at half intensity maximum (FWHM) of the spot is 0.611$\mu$m. (d) zoomed $\rho,z$ side view of the metalens. Thickness is 680nm, and the minimum design/fabrication resolution is 40nm in the radial direction with a material refractive index of 1.59. (e) Longitudinal ($\rho$,z) intensity profile where transmission efficiency and transmission normalized focusing efficiency are 87.46\%, 85.50\%, respectively. (f) Top view of the SEM image for a fabricated inverse-designed metalens. (g) Close-up image of the SEM, showing tiny bridges to support the high-aspect-ratio (680/40$\approx$17) nanoring-structure.}
\label{Inverse_design_lens}
\end{figure*}

We designed metalenses for the ZPAL system at 0.57 NA with 135 $\mu$m diameter (97 $\mu$m focal length). Figure~\ref{EMT_lens}(a) shows the optimized metalens with the EIM approach. It has a refractive index gradient varying from 1.00 to 1.59. The effective-medium-based metalens shows 81.63 \% efficiency at 97 $\mu$m focal plane while its transmission normalized focusing efficiency is 79.98 \%. Compared to the diffraction-limited focusing efficiency (total power within the third zero of the Airy pattern $\approx$ 93.8 \%~\cite{born2013principles}), our EIM-based metalens shows relatively high efficiency even with rough fabrication constraints (40nm $\times$ 680nm minimum feature size) as shown in Fig.~\ref{EMT_lens}(d). Figures~\ref{EMT_lens}(f, g) show a realized EIM-based metalens where we further simplify the metalens into 170 nm wide unit cells, resulting in a certain efficiency drop due to the change of the unit-cell width from 40nm to 170nm. Then, we convert grayscale refractive indices to a combination of rectangular holes (air) and surrounding materials (CSAR-62), i.e., EIM. Rectangular holes were programmed into each such cell, as illustrated in Fig.~\ref{EMT_lens}(f), with their azimuthal angle rotated in accordance with the cell azimuth. Rotational EIM-based metalenses were fabricated on borosilicate-glass samples (obtained from an online vendor) coated with 185 nm of Indium-tin-Oxide (ITO). The substrates were cleaned using UV ozone and an adhesion promoter spun on at 4000 RPM. 680-nm-thick CSAR resist was achieved at 1250 RPM. Substrates were baked on the hotplate at 150C for 2 min. Finally, the water-soluble conducting film was spun at 4000 RMP. Beamer proximity-effect-correction (PEC) software was used to pattern at 1nA beam current, 2.5 nm stepping distance, 125keV beam energy, and 15ns dwell time. This yields a base dose of 240$\mu$C/cm$^2$.

The fabricated metalens shows transmission normalized focusing efficiency of 48 \%. We assume that the conversion from 40-nm-width unit-cells to 170-nm causes a significant efficiency drop as expected in our previous work~\cite{chung2020high}. The fabrication imperfection were relatively mild compared to the inverse-designed metalens, discussed in the next section. As in many other publications~\cite{khorasaninejad2016polarization, khorasaninejad2016metalenses,shrestha2018broadband, chen2017gan}, it is hard to validate unit-cells-based metalens design in full-wave simulations due to a computational expense of a large area metalens. In the next section, we fully utilize both the accuracy of the full-wave simulation and the fast convergence of the inverse design algorithm to discover a `full-wave validated' inverse-designed metalens.

\subsection{Inverse-designed metalens}
In this subsection we optimize and fabricate an axisymmetric inverse-designed metalens for ZPAL. To avoid the efficiency drops on: (1) the unit-cell conversions (40-nm-width to 170-nm-width) and (2) gradient refractive indices to EIM structure conversions, we choose two material system (air and CSAR-62), and fabrication-compatible minimum feature size (40nm $\times$ 680nm). In this way, the result of the full-wave simulation does not suffer efficiency drops on such conversions. Fabrication perfection is required to maintain the simulated efficiency.

As shown in Figs.~\ref{Inverse_design_lens}(d), the inverse-designed axisymmetric metalens only has refractive index values of 1.0 and 1.59 at 405 nm wavelength. Figure~\ref{Inverse_design_lens}(e) shows longitudinal ($\rho$,z) intensity profile where transmission efficiency and transmission normalized focusing efficiency are 87.46\%, 85.50\%, respectively. Surprisingly, the transmission normalized focusing efficiency of our inverse-designed metalens exceeds that of the gradient refractive indices lens by 5.2\% with only a single material. We analyze the main reasons for high efficiency in two ways. First, unit-cell-based design approaches are relatively vulnerable to rapidly oscillating phase and amplitude profiles because the rapid change of phase/amplitude between neighboring unit cells may fail to create a desired continuous wavefront~\cite{chung2020high}. In the worst case scenario, only one or two unit cells are placed to model continuous variation of 2-$\pi$ phase shift in our 0.57 NA EIM-based metalens design. Second, the inverse design technique inherently converges to the local optimum quickly. Therefore, inverse design may often get stuck on bad local optima because of complicated optimization issues~\cite{park2022free}. However, our design problem, 0.6 NA metalens with 135 $\mu$m diameter, has relatively strict fabrication constraints leading to simple optimization problems in local areas. At the same time, it is still a large-area optimization with a total of 6750 design parameters. In other words, the design parameters are strongly correlated in local areas but nearly uncorrelated among largely-spaced parameters. Therefore, the summation of the local solutions found by our inverse design could be close to the upper limit.

To fabricate an inverse-designed metalens, a dose of 194 $\mu$C/cm$^2$ was applied, and supporting bridges were added to avoid collapsing the high-aspect-ratio nano-ring structure. However, despite nano-bridge support, some material adhered to the substrate surface while parts of the upper layers were torn away due to tensile stress. The measured transmission normalized focusing efficiency dropped to 36\%. It could be improved by changing the material to a high index material, which reduces the aspect ratio to cover 2$\pi$ phase range.

\section{Conclusion}
We designed a full-wave validated axisymmetric metalens for the ZPAL system via inverse design which shows greater focusing efficiency than gradient indices theoretical lenses. The normalized transmission efficiency of the optimized metalens (85.50 \%) is close to a theoretical upper limit for the given dimension, meaning that axisymmetric metalens designs can have higher efficiency than conventional convex microlens and unit-cell designed metalens. ZPAL with metalens-array implementation may open a new way of fabricating micro/nano-pattern at a faster and lower cost compared to nano-patterning electron beam lithography which often suffers high-cost and slow writing time. Furthermore, our design approach can be extended to much larger metalens designs since both the full-wave Maxwell solver (Meep)~\cite{oskooi2010meep} and adjoint optimization algorithm used in this work scale well with an increasing number of computing cores. Emerging fast-solver techniques such as integral equation-based Maxwell solver combined with adjoint optimization can also scale this approach to macroscopic length scales. Adjoint optimization itself has already proven its scalability in aerodynamic design~\cite{Aage2017}, where more than a billion voxels were simulated to find optimal distributions.

% \begin{acknowledgement}
%  The e-beam lithography was done by Mark Mondol at MIT. The computational resource was provided by KISTI (KSC-2022-CRE-0263).
% \end{acknowledgement}

% \begin{funding}
% This work was supported by the National Research Foundation of Korea (NRF) funded by the Ministry of Science and ICT (MSIT) (NRF-2022K1A3A1A91093776) and also under the High-Potential Individuals Global Training Program (IITP-2022-0-0154896), and Artificial Intelligence Graduate School Program (No.2020-0-01373, Hanyang University) supervised by the IITP(Institute for Information and Communications Technology Planning and Evaluation). 
% \end{funding}

\section{conflict of interest statement}
The authors declare no conflicts of interest regarding this article.

\appendix

% \section{\label{sec:level1}Fabrication imperfection (Feng or Hank)}

% \begin{figure*}[ht]
% \centering
% \includegraphics[width=1.0\linewidth]{EMT_fab_mid_zone.png}
% \caption{\hl{comment: I'm not sure we want to include this figure}Scanning-electron micrograph of the cleavage through mid zones of effective-index-modulated (EIM) Fresnel microlenses.}
% \label{Inverse_design_lens1}
% \end{figure*}

% \begin{figure*}[ht]
% \centering
% \includegraphics[width=1.0\linewidth]{Inverse_design_slice.png}
% \caption{\hl{comment: I'm not sure we want to include this figure}Cross sectional view of inverse-designed metalens, showing that 
% development occurred all the way down to the substrate surface.}
% \label{Inverse_design_lens2}
% \end{figure*}
% \newpage
% \section{\label{sec:level1}Extension to non-symmetric 3D metalenses}

% \begin{figure*}[ht]
% \centering
% \includegraphics[width=1.0\linewidth]{3D_metalens_20um_rev1.png}
% \caption{Inverse designed metalens in 3D. It takes approximately 28 hours to be converged in the 100 computational cores (one iteration (forward + backward simulations) of inverse design in 3D takes approximately 504 seconds while a cylindrical symmetry design with the same size only takes 4 seconds.}
% \label{Inverse_design_lens3}
% \end{figure*}
% \newpage
\section{Robustness of the optimized metalens}

\begin{figure*}
\centering
\includegraphics[width=1.0\linewidth]{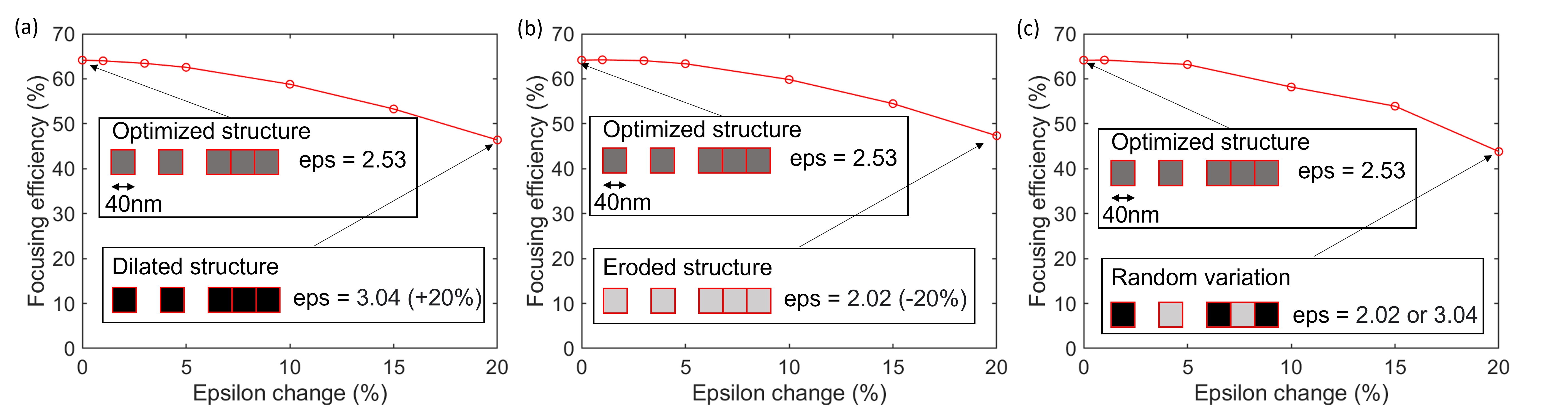}
\caption{Robustness test with (a) dilated structures, (b) eroded structures, and (c) random variations in the optimized metalens shown in Fig~\ref{Inverse_design_lens}. The permittivities of Yee's pixels were modified to introduce relatively small structural variation. The permittivity change effectively represents eroded or dilated structures via subpixel smoothing effect~\cite{oskooi2010meep}. A ten percent change in the permittivity degrades the transmission normalized focusing efficiency by 5 to 10 percent. It implies that the optimized metalens maintains its efficiency over slight fabrication error.}
\label{Robustness_check}
\end{figure*}

The robustness of the optimized metalens was tested by adjusting the permittivity of the finite pixels. The fixed pixel size makes it relatively hard to model a slightly delated or eroded structure in the finite difference scheme. Therefore, we utilize grayscale permittivity to model slightly delated or eroded structures. This method was first used in the subpixel smoothing in the FDTD~\cite{oskooi2010meep} where interpolating the permittivity over the material surface can significantly improve the stair-case error of the FDTD. As shown in Fig.~\ref{Robustness_check}, our optimized metalens has relatively robust for the slightly delated or eroded structures.

% \newpage
 \section{Optimized metalens data}
Geometry parameters for the inverse-designed metalens is listed in this section. Each grid occupies a nanoring of the 40-nm-width, 680-nm-height and the radius corresponding to $40\cdot\textrm{N}$ nm (N is the index of geometry parameter). The value “1” means permittivity of CSAR-64 at 405 nm, while the value “0” means air:

\noindent1 1 1 1 1 1 1 1 1 1 1 1 1 1 1 1 1 1 1 1 1 1 1 1 1 1 1 1 1 1 1 1 1 1 1 1 1 1 1 1 1 1 1 1 1 1 1 1 1 1 1 1 1 1 1 1 1 1 1 1 1 1 1 1 1 1 0 0 1 1 0 0 0 0 0 1 1 0 0 0 0 0 0 1 0 0 1 0 0 0 0 1 1 0 0 0 1 0 0 0 0 0 0 0 0 0 0 0 0 0 0 0 0 0 0 0 0 0 0 0 0 0 0 0 0 0 0 0 0 0 0 0 0 0 0 0 0 0 0 0 0 0 0 0 0 0 0 0 0 0 0 0 0 0 0 0 0 0 0 0 0 0 0 0 0 0 0 0 0 0 0 0 1 1 1 1 1 1 1 1 1 1 1 1 1 1 1 1 1 1 1 1 1 1 1 1 1 1 1 1 1 1 1 1 0 1 1 1 1 0 1 1 0 1 1 1 1 1 0 0 1 1 1 0 0 1 1 0 1 0 1 1 1 0 0 1 1 1 1 0 0 0 0 0 1 1 0 1 1 0 0 0 0 0 0 0 1 1 1 0 0 0 0 0 0 0 0 1 0 0 1 0 0 0 0 0 0 0 0 0 1 1 1 1 1 1 1 1 1 1 0 1 1 1 1 1 1 0 1 1 0 1 1 1 0 1 1 1 0 0 1 1 1 0 1 1 0 1 1 1 0 0 0 1 1 1 1 0 0 0 0 0 1 0 1 1 0 0 0 0 0 0 0 1 1 0 1 0 0 0 0 0 0 0 0 1 1 1 1 1 1 1 1 0 0 0 1 1 1 1 1 1 1 0 0 1 0 1 1 1 1 0 0 0 1 1 1 1 0 0 0 1 1 1 0 0 0 0 0 1 1 0 1 0 0 0 0 0 1 0 1 1 0 0 0 0 0 1 1 1 1 1 1 1 1 0 0 1 1 1 1 1 1 1 0 0 1 1 1 1 0 1 0 1 1 1 1 0 0 1 0 1 1 0 0 0 0 1 1 0 0 1 0 0 0 1 1 0 0 0 0 0 1 1 1 1 1 1 1 1 0 1 1 1 1 1 0 1 1 1 0 1 0 1 1 0 1 1 1 0 0 0 0 1 1 0 0 1 0 1 0 0 0 1 0 0 1 0 0 0 0 1 1 1 1 1 1 1 0 1 1 1 1 1 1 0 1 1 0 0 1 1 1 1 0 0 1 0 0 1 1 1 0 0 0 0 0 0 1 1 0 0 0 0 0 0 1 1 1 1 1 1 1 1 0 1 0 1 1 1 1 0 1 0 1 1 1 0 0 1 1 1 0 1 0 0 0 1 1 0 0 0 0 1 0 0 0 0 1 1 1 1 1 1 1 1 0 1 1 0 1 1 1 1 0 1 1 0 0 1 1 1 0 0 1 0 0 0 0 1 1 0 0 0 0 0 0 1 1 1 1 1 1 1 0 0 1 1 1 1 0 1 1 0 0 0 0 1 1 0 1 0 0 0 0 0 1 0 1 1 0 0 0 0 1 1 1 1 1 1 1 0 0 1 1 1 1 1 0 0 0 1 1 1 0 1 1 0 0 0 0 0 1 1 0 0 0 0 0 1 1 1 1 1 1 1 0 1 1 0 1 1 1 1 0 1 0 1 1 1 0 0 0 0 0 1 1 0 0 1 0 0 0 0 1 1 1 1 1 1 1 0 1 1 0 1 1 1 0 1 0 1 1 1 0 0 0 0 1 0 1 1 0 0 0 0 1 1 1 1 1 1 1 0 1 1 1 0 1 1 1 0 0 0 1 1 0 1 0 1 0 0 0 1 0 0 0 1 1 1 1 1 1 1 0 0 1 1 1 1 0 0 0 1 1 0 1 0 0 0 0 0 1 1 0 0 0 1 1 1 1 1 1 1 0 0 0 1 1 1 1 0 0 1 1 1 0 0 0 0 0 1 1 0 0 0 1 1 1 1 1 1 1 0 0 1 1 0 1 1 0 0 1 1 0 0 0 1 1 0 0 0 0 0 1 1 1 1 1 1 1 0 0 1 1 1 0 1 0 0 0 1 0 0 1 1 0 0 0 0 0 0 1 1 1 1 1 1 1 0 1 1 1 0 1 0 1 1 0 0 1 0 1 1 0 0 0 0 0 1 1 1 1 1 1 0 0 1 1 1 0 1 0 0 1 1 0 0 1 1 0 0 0 0 0 1 1 1 1 1 1 1 0 1 0 1 1 0 0 1 1 1 0 0 0 1 1 0 0 0 0 1 1 1 1 1 1 0 0 1 1 1 0 1 1 0 1 0 0 0 1 1 0 0 0 0 1 1 1 1 1 1 0 0 1 1 1 1 0 0 1 0 1 0 1 0 1 0 0 0 1 1 1 1 1 1 0 0 1 1 1 1 0 0 0 0 1 1 1 0 0 0 0 0 0 1 1 1 1 1 1 0 1 0 1 1 0 1 1 0 0 0 1 0 1 0 0 0 0 1 1 1 1 1 1 0 1 0 1 1 1 0 0 1 0 0 1 1 0 0 0 0 1 1 1 1 1 1 0 0 1 1 0 1 0 0 0 1 0 1 0 0 0 0 1 1 1 1 1 1 0 0 1 1 1 0 0 0 0 1 1 0 1 0 0 0 0 1 1 1 1 1 1 0 0 1 1 1 0 0 1 1 0 0 0 1 0 0 0 1 1 1 1 1 1 0 0 1 1 1 0 0 0 0 1 1 0 0 0 0 0 1 1 1 1 1 1 0 0 1 1 1 0 0 1 1 0 1 0 0 0 0 1 1 1 1 1 1 0 1 1 1 0 0 0 0 1 1 0 0 0 0 0 1 1 1 1 1 1 0 1 1 1 0 0 0 0 1 1 1 0 0 0 0 1 1 1 1 1 1 0 1 1 1 0 0 1 0 1 0 0 0 0 0 0 1 1 1 1 1 1 0 1 1 1 0 1 0 1 0 0 0 1 0 0 1 1 1 1 1 1 0 1 1 1 0 0 1 0 1 0 0 1 0 0 1 1 1 1 1 1 0 1 1 1 0 0 0 1 1 0 0 1 0 0 1 1 1 1 1 1 0 1 1 1 0 1 0 1 0 1 0 0 0 0 1 1 1 1 1 0 1 0 1 1 0 1 1 0 0 0 0 0 0 0 1 1 1 1 1 0 0 1 1 0 1 1 0 0 0 1 0 0 0 0 1 1 1 1 1 0 0 0 1 1 1 0 0 0 0 0 0 0 0 1 1 1 1 1 0 0 0 1 1 1 0 0 0 0 0 1 0 0 1 1 1 1 1 0 0 1 0 1 1 0 0 0 0 0 1 1 1 1 1 1 0 1 1 1 0 0 0 1 1 0 0 0 0 0 0 1 1 1 1 1 0 0 0 1 1 1 0 0 0 0 0 0 1 1 1 1 1 0 1 1 0 0 1 1 0 0 0 0 0 0 1 1 1 1 1 1 0 0 0 1 1 1 0 0 0 0 0 0 1 1 1 1 1 1 0 0 0 1 1 1 0 0 0 0 0 0 0 1 1 1 1 1 0 0 0 1 1 1 0 0 0 0 0 0 0 1 1 1 1 1 0 0 0 1 1 1 0 0 0 0 0 1 1 1 1 1 1 0 0 1 0 1 1 0 0 0 0 0 0 1 1 1 1 1 1 0 1 0 1 1 0 0 0 0 0 0 1 1 1 1 1 1 0 1 0 1 1 0 0 0 0 0 1 1 1 1 1 1 0 1 0 1 1 1 0 0 0 0 0 0 0 0 1 1 1 1 1 0 0 1 0 1 1

% Bibliography
%\bibliography{apssamp}
%apsrev4-2.bst 2019-01-14 (MD) hand-edited version of apsrev4-1.bst
%Control: key (0)
%Control: author (8) initials jnrlst
%Control: editor formatted (1) identically to author
%Control: production of article title (0) allowed
%Control: page (0) single
%Control: year (1) truncated
%Control: production of eprint (0) enabled
\providecommand{\noopsort}[1]{}\providecommand{\singleletter}[1]{#1}%

\end{document}